# Transitions in
# Oscillatory Dynamics of Two Connected Neurons with Excitatory Synapses

## Yasser Roudi[*], Shahin Rouhani

Department of Physics, Sharif University of Technology, Tehran, Iran
and
School of Intelligent Systems Institute for Studies in Theoretical Physics and Mathematics, Tehran,
P.O. Box 19395-5746, Iran

## Abstract

It is shown that long term behavior of two connected Integrate- and- Fire neurons with excitatory synapses is determined by some fixed-points. In the case of equal synaptic weights four different dynamic phases are found. Between these phases there is a specific phase with a global attractor fixed-point, which is of interest from different viewpoints. Simulations support our analytic work. When synaptic weights are equal we observe no synchronization but with different weight we do observe an almost synchronous state. Simulations show that when there is only one non-trivial fixed-point the period of oscillations is stable against small changes in synaptic weights.

*Address from 13 March 2002:
Program in Neuroscience, SISSA, via Beirut 4, I- 34014
Trieste, Italy
yasser@sissa.it

# 1. Introduction

Neural oscillators play a crucial role in modeling of brain function. As the most general pattern generator network, neural oscillators appear in various parts of nervous system, above all in rhythmic sensory motor activity (Marder and Calabrese 1996; Abbott and Marder 1998) and sensory processing (e.g. olfactory bulb and auditory processing) (Li 1995; Li and Hoppfield 1989; Leen et al. 1991; Yao and Freeman 1989; Brown and Cooke 1998, 1995). As a general rule, oscillator circuits have a vital role in every electronic system: synchronization of different parts of the system cannot be done without them. Neural oscillators may in fact play a similar role. In addition to neural modeling, such oscillators have been considered in neural circuits for construction of sensory-motor systems of robots (Kimura et al 1999).

The simplest neural circuit, which can show oscillatory behavior, is a single neuron with a continuous excitatory input (Bear 1998; Izhikevich 2000). These oscillatory behaviors result from the membrane ionic mechanism, which in some way can be included in biological models of single neurons (e.g. Hodgkin-Huxley model). When one considers simpler models like Integrate-and-Fire neurons, this oscillatory response can be seen, because IFN would reset after each spike. Resetting the neuron is an unrealistic and an artificial supposition and when one replaces this behavior with a decaying membrane potential, oscillatory behavior cannot be seen unless assuming an absolute refractory time.

Here we study the behavior of the next simplest oscillating neural system, two neurons linked with excitatory synapses. In this manuscript, ignoring lots of details in the dynamic of neurons, e.g. specific dynamics of ion channels and action generation we use a modified version of Leaky Integrate-and-Fire neuron both in analytical studies and simulations. IFN model is developed by Lapicque (Lapicque 1907) and replicates a neuron with an electrical circuit consisting of a parallel capacitor and resistor. Our model differs from standard IFN by the statement that instead of resetting the neuron, we consider a more realistic assumption that membrane potential would decay exponentially following each spike. To begin with, the input current is chosen strong enough so that one of the neurons fires a spike, but the current will be removed at later times. Thus we are dealing with an initial condition such that one of the neurons is set at the firing threshold and the other one at some smaller membrane potential. Other kinds of synaptic connections as well as presence of external currents have been considered by other authors (Herrera et al 1997;Luo and Harris 2001; Tino et al 1995). In these cases the system may show stable oscillatory response, as in (Herrera et al 1997;Luo and Harris 2001) through an electronic analog circuit simulation. Using geometrical reasoning and discrete-time dynamics stable oscillatory response was studied in (Tino et al 1995). Dynamical behavior using simplified model of a single neuron in networks, which differ from ours in some respects, is also studied in (Gomez and Budelli 1996). And in (Bose et al 2000), the coupled oscillator dynamics of model neurons is studied with emphasis on synchronous states of the system where they show that almost synchronous states exist However their model neurons are not IFNs.

## 2. Single Neuron Model

We use the "Integrate & Fire " model (Koch 1999; Lapicque 1907; Tuckwell 1988) for a single neuron through out this work, with the additional assumption discussed above. Although Integrate-and-Fire model is more biological in comparison with simple binary neurons, it dose not deal with detailed biological facts in dynamics of ion channels and other membrane mechanisms. These mechanisms are considered for example in Hodgkin-Huxley model (Hodgkin and Huxley 1959), which is based on Sodium ion flow, Potassium ion flow and leakage ion flow. The simplifications of IFN make it an appropriate model for studying different aspects of neuronal dynamics in comparison to realistic models like Hodgkin-Huxley neurons. This fact motivated various authors to use this model in the simulation or analytic study of neural basis of brain function especially when one deals with large networks (Amit and Brunel 1997; Battaglia and Treves 1998).

In this model the state of neuron is determined by membrane potential that follows (1) bellow:

$$\frac{dV}{dt} + \frac{V}{\tau} = I(t),\qquad(1)$$

where $V$ is the membrane potential of soma, $\tau$ membrane time constant (a measure of the time taken to decay due to the "leaky" nature of the cell membrane) and $I(t)$ the input current to the neuron. Whenever the membrane potential of the neuron reaches a threshold of $\theta$ it will fire which means that it sends out an action potentials to other neurons connected to its axon and then its membrane potential will decay (here as an exponential function). The action potential can be identified by a response function $p(t)$. Usually after reaching the threshold out put potential grows up to a maximum and then falls off. Here we consider an alpha-function (2) as the response function.

$$p(t) = \begin{cases} te^{-\alpha t} & t \geq 0 \\ 0 & t < 0 \end{cases} \qquad . \qquad (2)$$

There are more biological plausible forms for response function at different levels of abstraction such as those found in (Koch 1999), but we use above for simplicity. In general the current $I_i(t)$ in to $i^{th}$ neuron at time t can be split as:

$$I_i(t) = I_i^{ext}(t) + I_i^{int}(t), \qquad (3)$$

where $I_i^{int}$ and $I_i^{ext}$ are respectively the current received from the firing neurons of network and the current from external sources.

From above, if $t_i^k$ represents the time of $k^{th}$ firing of the $i^{th}$ neuron the internal input current to the $j^{th}$ neuron $I_j^{int}(t)$ is:

$$I_j^{int}(t) = \sum_{i,k} w_{ji}\, p(t - t_i^k), \qquad (4)$$

In fact the internal current to $j^{th}$ neuron $I_j^{int}(t)$ obeys the equation (5) below (Amit and Brunel 1997):

$$\frac{dI_j^{int}}{dt} + \frac{I_j^{int}}{\tau_c} = \frac{\tau}{\tau_c} \sum_{i,k} w_{ji}\, p(t - t_i^k), \qquad (5)$$

but when the time constant of synaptic conductance change $\tau_c$, is considerably smaller than the membrane time constant $\tau$, this equation can be approximated by (4).

We have considered the same response function for each neuron in the network.

## 3.The Two-Neuron Oscillator

Consider the structure consisting of two coupled Integrate-and -Fire neurons with excitatory connecting synapses. Suppose that membrane potential of the first is equal to threshold $\theta$ at $t=0$ and the second one is below threshold .So the first neuron will send an action potential to second one at this time and its membrane potential will decay exponentially. On the other hand the current to the second neuron would cause its potential to vary with time according to (1). Therefore it first grows up, and if it reaches threshold, it will fire, otherwise after reaching a maximum it will fall off. Explicitly, the functional form of membrane potential of the second neuron before reaching a threshold is as follows:

$$V_2(t) = W_{21}e^{-ct}(1/b^2 - e^{-bt}(t/b + 1/b^2)) + e^{-ct}V_o \,, (6)$$

where 'c' is the inverse of the membrane integration time constant and $b = \alpha - c$. Undoubtedly occurrence of each state depends on the values of parameters; in some ranges the first case and in the other ones the second will occur. If the second neuron cannot reach threshold the potential of both neurons will fall to zero eventually. But if it can attain the firing threshold, the situation repeats for the first neuron, then one of the two cases can happen again. What is the long-term behavior of two neurons in the non-trivial case?

## 4. Fixed Points

Think about the state of system, where the first neuron is at firing threshold and the second one has a membrane voltage of $V_2^0 \neq 0$ (calling it a *C-state*). We can easily find the state of the system (the membrane voltage of second neuron) at the previous firing time of first neuron ( *P-state*, If any!). It is clear that before arriving at the state under consideration there has been a situation at which the second neuron has fired (*M-state*). If we consider the exponential decay form of membrane potential after firing as (7) the time difference between the *M-state* and *C-state* is just (8):

$$Decay(t) = \theta e^{-\gamma t} \,, \qquad\qquad (7)$$

$$t = Ln(\theta / V_2^0)\big/ \gamma \,. \qquad\qquad (8)$$

So voltage of the first neuron at *M-state* will be:

$$V_1^m = \theta e^{ct} - W_{12}(1/b^2 - e^{-bt}(t/b + 1/b^2)) \ . \ (9)$$

Repeating the sequence of *M-state* and *P-state* one can find the expressions below for the membrane potential of second neuron at *P-stat:*

$$t' = Ln(\theta / V_1^m)\big/ \gamma \ . \qquad\qquad (10)$$

$$V_2^{0'} = \theta e^{ct'} - W_{21}(1/b^2 - e^{-bt'}(t'/b + 1/b^2)) \ . \ (11)$$

 It is obvious from (7-10) that the previous state can be determined explicitly for a *C-state*. Also it is clear that if either $V_1^m$ or $V_2^m$ were negative or grater than $\theta$, there would be no *P-state* corresponding to the *C-state* under consideration. Actually for these situations there would be no state from which, after passing some middle states one can reach it and so they can be only the initial state of the system. There is no initial state that can include these states in its pathway through phase space of the system.

 In order to find fixed-points we first must consider to the property that distinguishes these points from the other points in the phase diagram, which is the fact that if we put the system initially at a fixed-point, it remains there for ever.

One trivial fixed-point is obviously $V_2^0 = \theta$, the state of successive rapid firings of neurons. The other fixed point is the silent state that none of the neurons fire. Other fixed-points must satisfy $V_2^{0'} = V_2^0$ or:

$$V_f^m = \theta(\theta/V_f)^{c/\gamma} - W_{12}(1/b^2 - (\theta/V_f)^{-b/\gamma} (\ln(\theta/V_f)/(\gamma b) + 1/b^2))$$

$$V_f = \theta(\theta/V_f^m)^{c/\gamma} - W_{21}(1/b^2 - (\theta/V_f^m)^{-b/\gamma} (\ln(\theta/V_f^m)/(\gamma b) + 1/b^2)) \quad (12)$$

It should be noted that these nonzero fixed-points are oscillatory fixed-points with a period of $T = t + t'$.

Solving these equations analytically is hard, but we can check some properties of the fixed-points, such as number of them and the kind of each one (attractive or repulsive) by analyzing the behavior of the equation above. We do this, for the case of equal synaptic weights. In this case some general properties of the long-term behavior of the system can be established. This analysis can be done for the case of different synaptic weights in a similar manner, but finding general properties are rather a hard work.

## 4.1. Special Case of Equal synaptic weights

The reader can convince himself that when synaptic weights are equal, the symmetry of equations causes that at the fixed-points, the minimum membrane potential of neurons in each cycle would be equal. Thus in this case, (12) reduces to (13):

$$f(V) = \theta(\theta/V)^{c/\gamma} - W(1/b^2 - (\theta/V)^{-b/\gamma} (\ln(\theta/V)/(\gamma b) + 1/b^2)) \quad (13)$$

$$V_f = f(V_f)$$

So we can determine fixed-points, finding the intersection points of two graphs, $y_1 = V$, $y_2 = f(V)$ which satisfy the condition that $0 \le V \le \theta$.

After all, one must considers a critical point, which is necessary for following arguments. As a biological fact, for most neurons experimental results show that $\tau \sim 10-20 ms$ and $1/\alpha \sim 2-3 ms$, and so $b=\alpha-c>0$. This ensures us about the unbounded growing of membrane voltage of neuron after receiving a spike, which clearly is not biologically plausible.

We begin analysis of (13) by finding maximum and minimums of $f(V)$. Setting the derivative equal to zero leads to:

$$V^{-\alpha/\gamma} = \frac{w\theta^{\left(\frac{\alpha}{\gamma}-1\right)}}{c\gamma}\ln\left(\frac{\theta}{V}\right) \qquad (14)$$

It is clear that (14) has two solutions, both of them smaller than threshold. In addition it is obvious from this fact, as well as limiting behavior of $f(V)$, that the smaller one is a local minimum of $f$ and the largest one a local maximum. So $f$ has a typical shape as shown in Fig.1.

In Fig.2 four different relative positions of $y_1=V$ and $y_2=f(V)$ are shown. Each one of these diagrams corresponds to a different phase in the dynamic of the system, as we are going to describe below. From now on we do not take attention to the trivial fixed point $V_f=\theta$ because it exists in the same manner in all phases (every where it is a repulsive fixed-point):

**a)** Fig2.a corresponds to the dynamical phase, in which there are only two trivial fixed points. The larger one is repulsive and the smaller one a global attractor.

**b)** In the case of Fig2.b one can distinguish, an additional attractor ($V_{att}$) which does not exist in phase (a) above as well as repulsive one ($V_{rep}$). Here the basin of attraction of the trivial attractive fixed point consists of initial voltages which lie between zero and $V_{rep}$ and the non-trivial attractor has the basin of attraction of $V_{rep} < V < \theta$. This reality can be understood easily, following the dynamic of each initial condition: Consider the problem of finding the ulterior state to a *C-state*. Solving the problem by itself is hard, but using the procedure described before for determining the *P-state* in reverse direction make it very easy. The only point to care is that there can be more than one such point for a definite *C-state* voltage. Which of them is the correct one? Every thing is deterministic, so it is expected that there would be only one determined state ulterior to each selected state. It is straightforward to see that the larger is the proper one. Remembering the shape of α-function it is clear that there are two solutions for the equation $V(t)=\theta$ but the appropriate one is the smaller. The soma potential of the other neuron at this time is the grater one.

**c)** In Fig2.c there are again two cross sections but using the argument above one finds that only the larger one can be accepted as a fixed-point. So in this phase there are two attractors, where the basin of attraction of the trivial fixed-point is $V_{\min} \leq V < \theta$ whereas the non-trivial attractor has the basin of attraction of $0 \leq V < V_{\min}$.

**d)** The last dynamic is nothing more than the foregoing case except that in this case the minimum of $f(V)$ reaches zero and so the basin of attraction of the trivial fixed point consists of it alone. So for all of initial condition the system finally reaches a unique oscillatory fixed-point.

The phase transition between a and b occurs when $y=x$ is tangent to $y=f(V)$. The system undergoes a transitions between b and c when one of the intersections of $y=x$ and $y=f(V)$ is located at the minimum of $f$. Finally the transition between c and d occurs when the minimum of $f$ reaches zero.

The last phase is of interest because each initial voltage leads to a determined and unique oscillatory dynamics, which clearly is a necessary condition for each oscillation generator circuit. The other property, which makes the two-neuron oscillator in the last phase as a suitable circuit for playing various role is the stability of the frequency of attractor against small changes in synaptic weights. We have examined this property through out some computer simulations. Results are shown in Fig.4. It is seen in this figure that this stability would be more powerful for stronger synapses as well as smaller decay constants. This kind of dependence to decay constant ($\gamma$) is reasonable. In fact we anticipate that when $\gamma$ reaches infinity (resetting the neuron after each spike) the dependence on synaptic weights would be more powerful. As well, when $\gamma=0$ no dependence on weights is expected. These two limiting behaviors agree with out simulation results.

## 5. Simulations

In order to support our analytical solutions, we have simulated the dynamic of the system using a C++ code with high numerical accuracy. Outcomes of this simulation are shown in Fig.3. As can be seen in these graphs one can distinguish different dynamics of the system as described above. Also we have used this simulation to discuss the synchronization of neurons with both equal and different synaptic weights in the following section. Also we have studied the dynamic of two-neuron network when synaptic weights are different using this simulation. We found that transitions similar to the equal synaptic weights can be seen in this case (Fig.3.d). In all of this simulations we have considered c=0.1 ms$^{-1}$ and b=0.36 ms$^{-1}$.

# 6.Oscillation Frequency and Synchronization

In order to study synchronization three time constants are important: the first and second are the time interval between successive firing of each neuron ($t$ and $t'$ in (8) and (10)) and the third one, the sum of two preceding time constants which obviously is the period of oscillation. "Full-Synchronization" occur when either $t$ or $t'$ are zero. This case does not occur unless one considers some unrealistic assumptions such as a delta function response of the presynaptic neuron that makes the postsynaptic neuron fire at the moment that presynaptic one fires. So in reasonable cases which some kinds of delay exists, the "Full-Synchronization" cannot be seen in this structure. Another interesting dynamics is the case of "almost-synchronous" dynamic which one of the time intervals between successive firing of neurons ($t$ or $t'$) are much less than the other one and consequently from the total period of oscillation e.g. $\dfrac{t'}{T} \ll 1$ (when $W_{21} > W_{12}$) or

$\dfrac{t}{T} \ll 0$ (when $W_{12} > W_{21}$).

When we studied equal weights, it was seen that the minimum voltage of both neurons through out oscillation phase are equal and so are times interval between firing of one neurons and the other one as is seen from (8) and (10). So, in this case the time interval between consecutive firings are comparable to the total period of oscillation of system ($T=t+t'$) and as a result one cannot observe any kind of synchronous dynamics (Fig 5.a). In contrast, when synaptic weights are different, voltage of neurons at fixed-point state are not similar, and there is a chance to see almost synchronous dynamics (Fig.5.b). In fact when synaptic weights are very asymmetrical, the synchronization between neurons increases as shown in Fig.6.Even in the case of different weights one can observe stability against small changes in synaptic strengths.

## 7. Discussion

In this manuscript we have studied the behavior of two neurons, linking with excitatory synapses, using Integrate-and-Fire model as a single neuron. We have focused on the case of equal synaptic weights because of its simplicity to analyze. We have found a dynamical phase of the system, which consist of only one attractor with the basin of attraction of whole initial state (i.e. membrane potential of each neuron). This phase is of interest because it shows that this neural circuit can act as an oscillation generator, with definite frequency, stable against initial membrane voltages of neurons. In addition to stability against initial voltages, it is seen throughout simulations that this attractor is stable against small changes in synaptic weights, which is another evidence for usefulness of this neural architecture oscillation generator.

Apart from these results, another importance of this study is its consequences in stability of autoassociators against small noises in the network. This kind of stability requires that inserting an excitatory current to a neuron in the network and making it fire, does not cause a resonance or synchronization with other neurons. Now think of a two-neuron sub network of the associative memory that one of them is at firing state results from noises of the network. This sub network cannot lie in phase (d), studied in the text, because in this phase the two-neuron sub network finally shows a resonance or synchronization between neurons that its consequences in the global dynamics of the network clearly oppose noise dying. Although this argument is restricted to pairs of cells, but its implication for autoassociative memories are linked to its extendibility to networks of interconnected cells, with couplings that may be pairwise weak and symmetric, but strong and asymmetric globally.

The result that increasing synaptic weights leads the system to the undesirable phase whether the synaptic weights are equal or different (phase (d) in symmetric synapses and almost synchronous dynamics in asymmetric case), is in agreement with the well-known fact that autoassociators perform well with weak couplings between pairs of neurons.

### Acknowledgment:

Authors would like to acknowledge Alessandro Treves and Valeria Del Prete for their precise and useful comments on this literature.

## Figure Legends:

**Fig 1:** A typical graph of $f(V)$ when b is positive

**Fig2:** Four different relative positions of $y_1 = V$, $y_2 = f(V)$, corresponding to four different dynamical phases of two-neuron network with equal synaptic weights.

**Fig 3:** $V_1$-$V_2$ (mV) phase portrait of the two connected neurons for different values of parameter corresponding to different phases discussed in the text. In a, b, c it is seen that changing $\gamma$ and $\theta$ for symmetric and constant weights, cause different dynamical phases. In each case (except a), non-trivial oscillatory fixed-point can be distinguished. Throughout d, the phase portrait for asymmetric synaptic weights is shown. Even in this case there is a global attractor. a) $W_{12} = W_{21} = 10^2 \text{mVms}^{-1}, \gamma = 0.2\text{ms}^{-1}, \theta = 40\text{mV}$. b) $W12 = W_{21} = 10^2 \text{ mVms}^1, \gamma = 0.1\text{ms}^{-1}, \theta = 40\text{mV}$, c) $W_{12} = W_{21} = 10^2 \text{ mVms}^{-1}, \gamma = 0.1 \text{ ms}^{-1}, \theta = 20\text{mV}$ d) $W_{12} = 10^2$, $W_{21} = 10^3 \text{mVms}^{-1}, \gamma = 0.1 \text{ ms}^{-1}, \theta = 40\text{mV}$. Starting points are selected from upper vertical line, corresponding to the state which first neuron fires.

**Fig 4:** Frequency of oscillation of global attractor in phase (d) vs. synaptic weights. Step like behavior of the graph shows stability of frequency against small changes in synaptic weights. Threshold is set equal to 20mV. Synaptic weights are in mVms$^{-1}$ and frequency in ms$^{-1}$. a) $\gamma = 0.1$ ms$^{-1}$, b) $\gamma = 0.4$ ms$^{-1}$

**Fig 5:** Simulation shows Non-Synchronous (a) and Almost-Synchronous (b) dynamics of two linked neurons. Corresponding values of parameters are a) $W12 = W_{21} = 80$ mVms$^{-1}$, b) $W_{12} = 1800, W_{21} = 80$ mVms$^{-1}$. In both graphs it is assumed that $\gamma = 0.1$ ms$^{-1}$, $\theta = 20$mV. As well t is in 10ms and potential in mV.

**Fig 6:** Ratio of t/T as a function of $W_{12}$ (mVms$^{-1}$) while keeping $W_{21} = 80$ mVms$^{-1}$. Using this graph it can be deduced that increasing the asymmetry between synapses leads to more synchronization between neurons. Also it is seen that even when synaptic weights are not equal some kind of stability against small changes in weights do exist. Other parameters are $\gamma = 0.1$ ms$^{-1}$, $\theta = 20$mV.

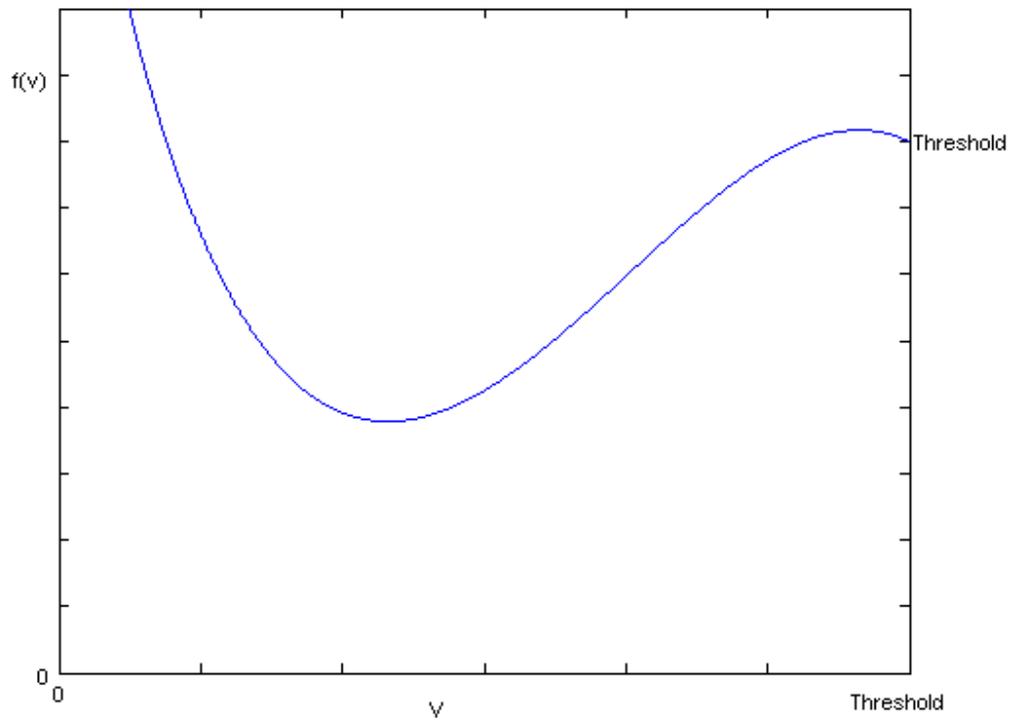

**Fig 1.**

]

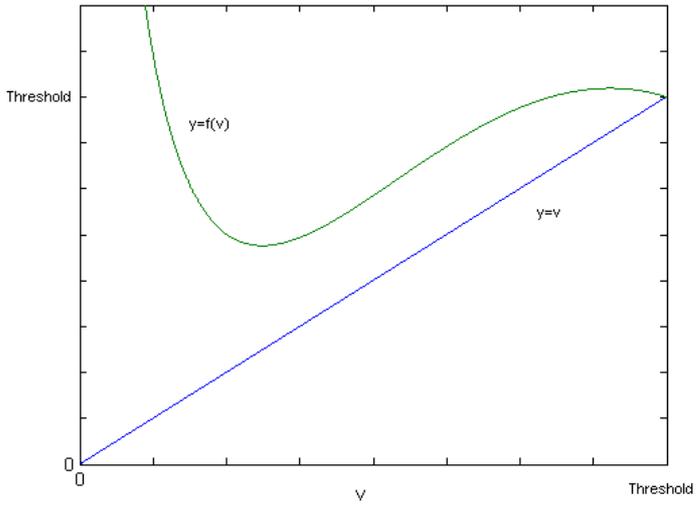

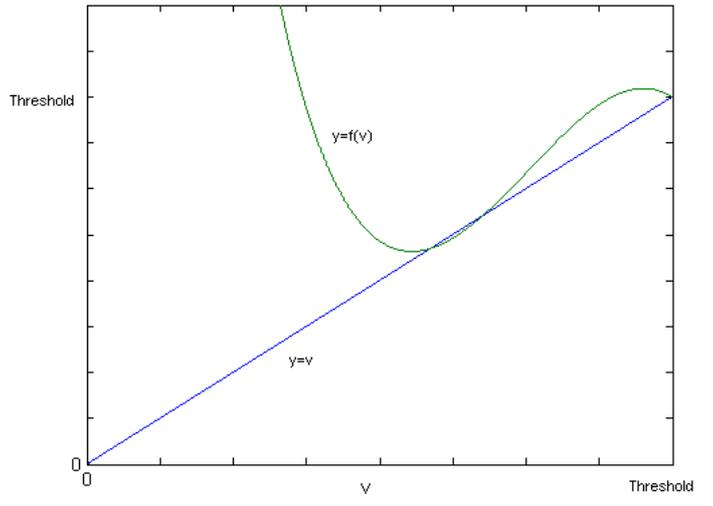

a.

b.

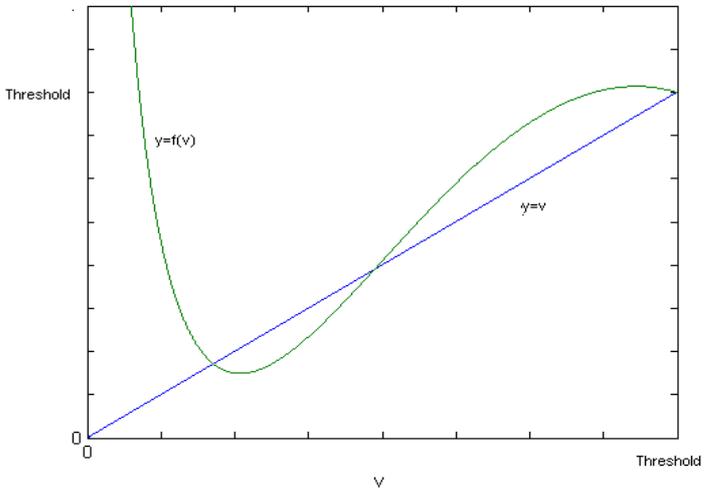

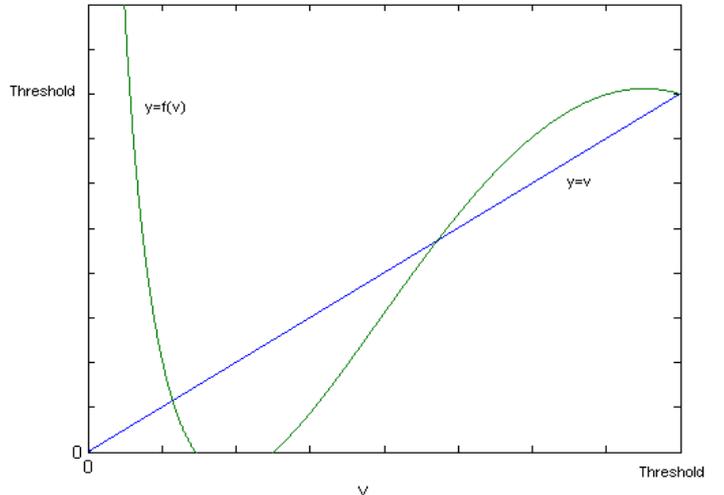

a.

b.

**Fig 2.**

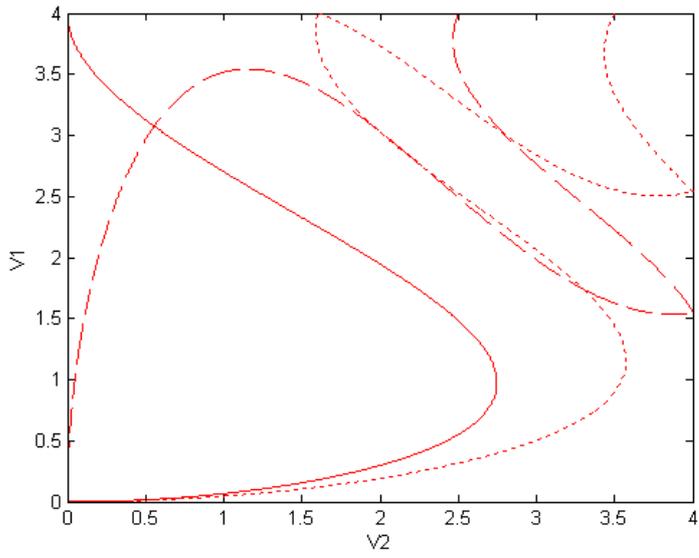

a.

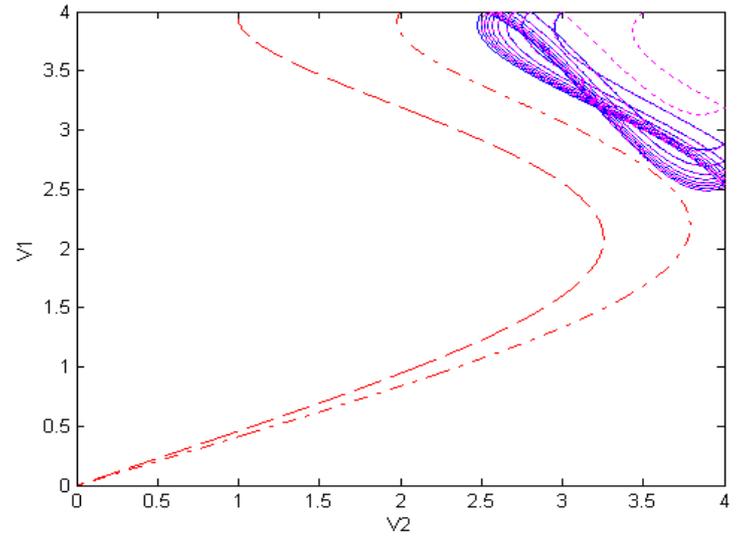

b.

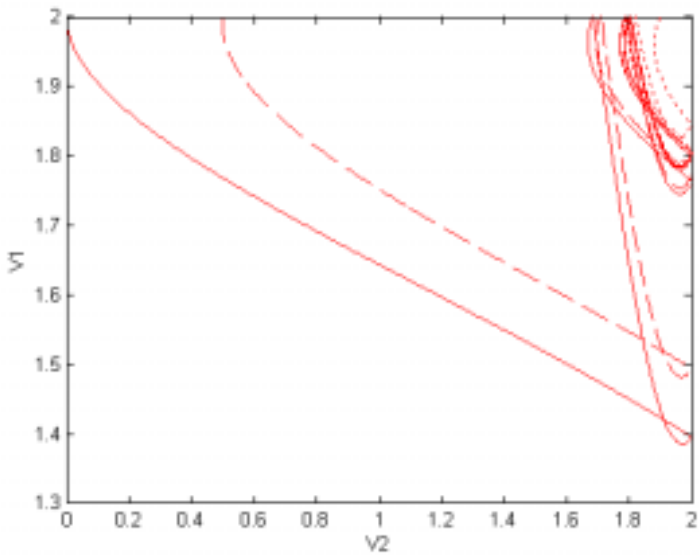

c.

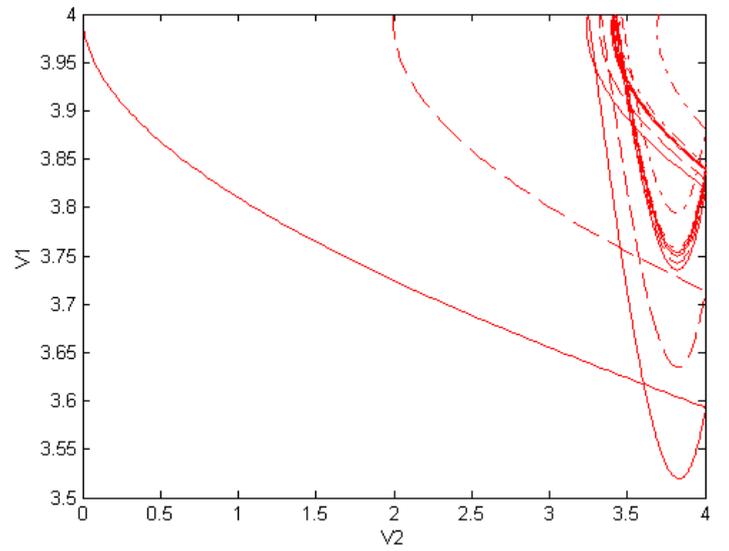

d.

**Fig 3.**

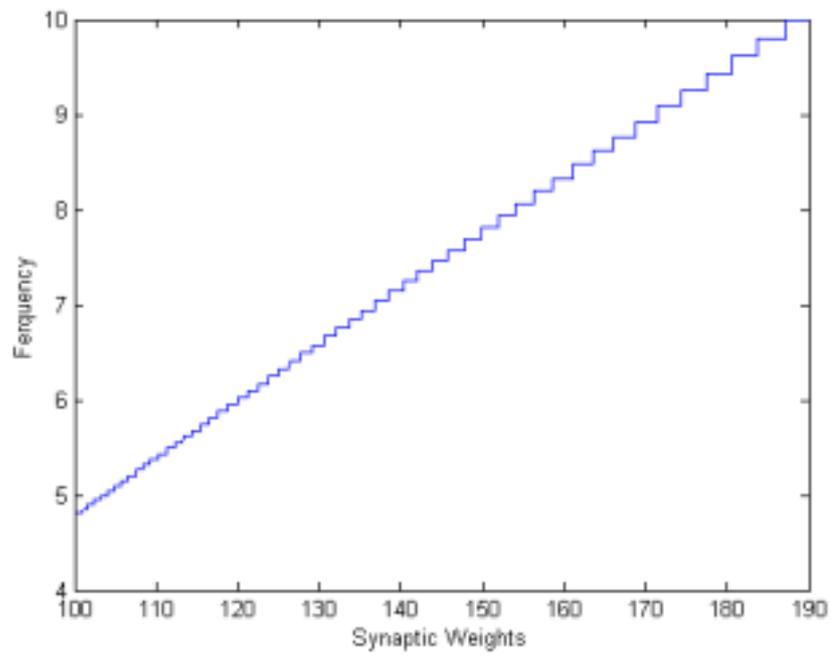

a.

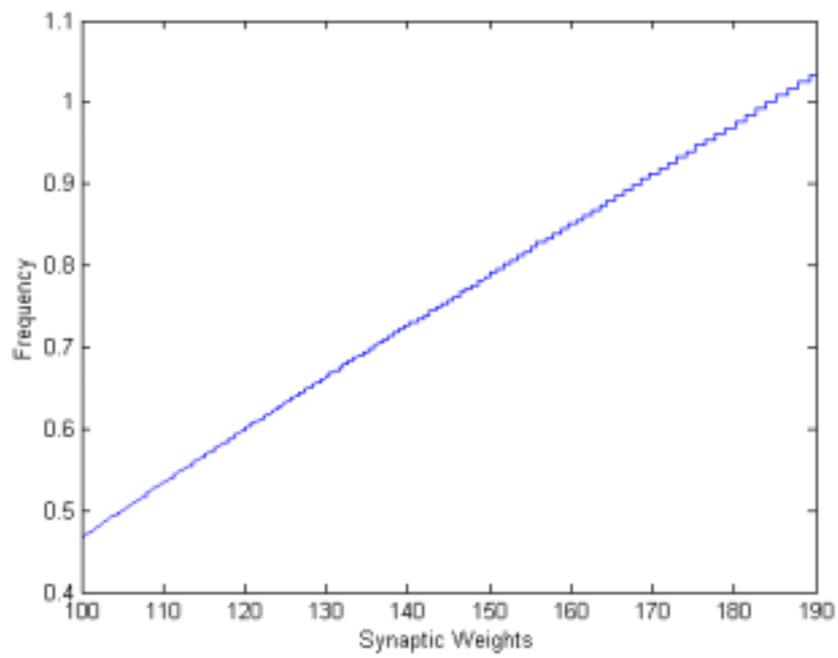

b.

**Fig 4.**

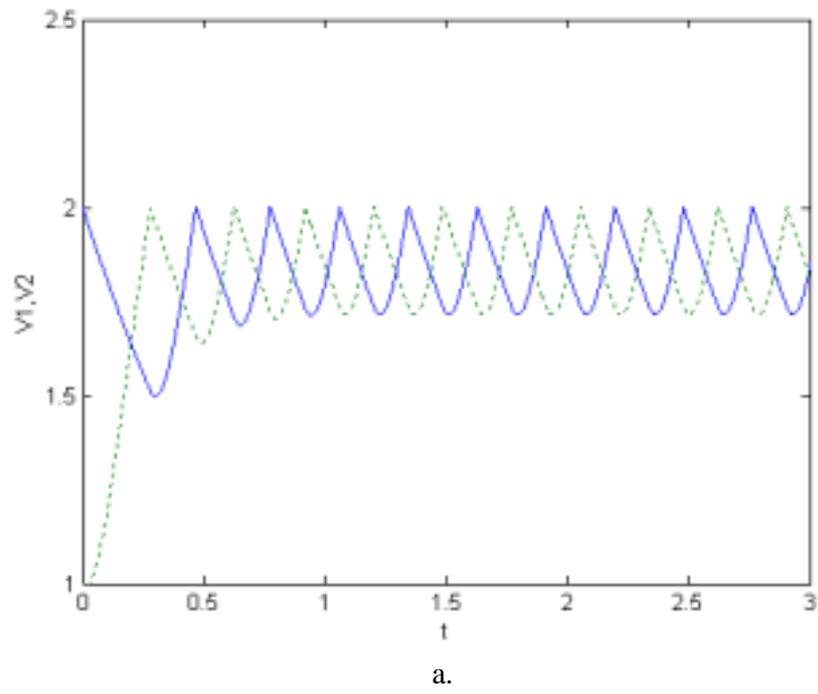

a.

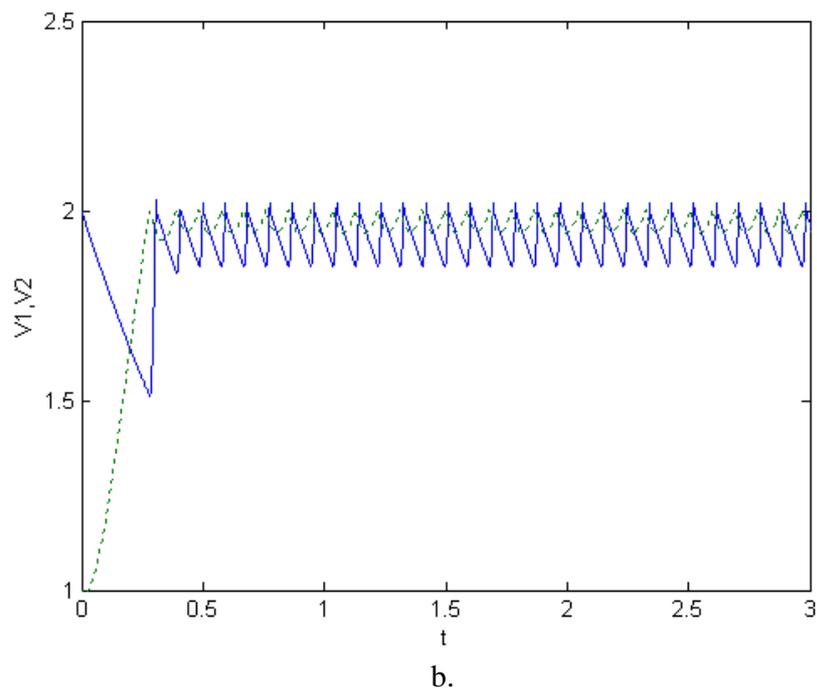

b.

**Fig 5.**

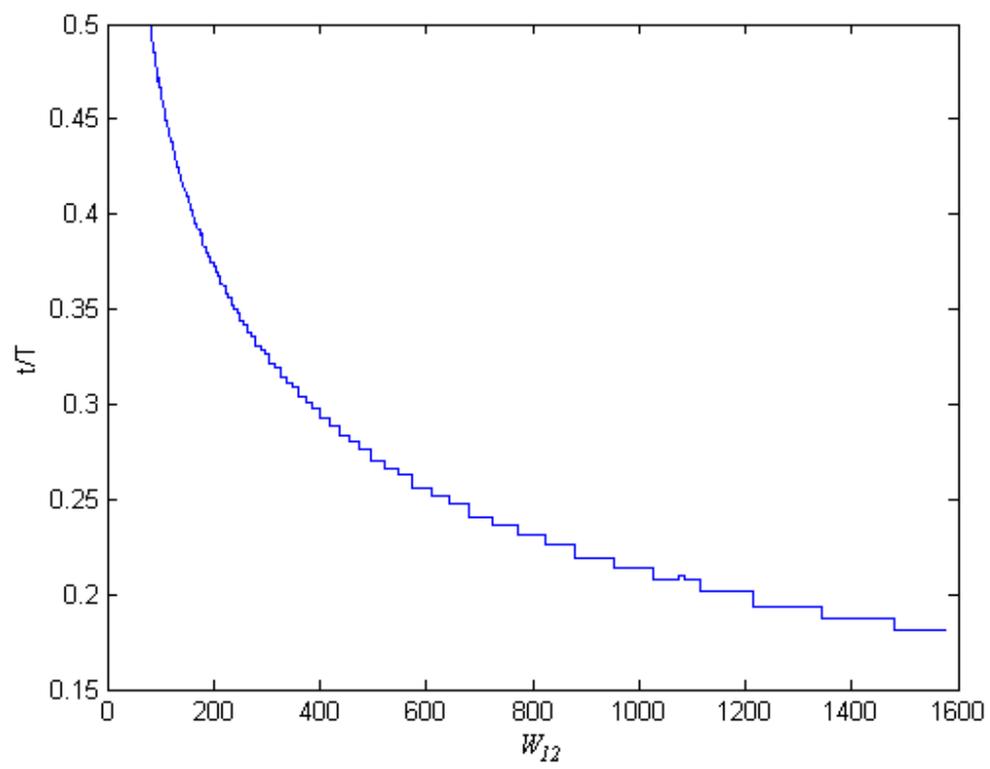

**Fig 6.**